\documentclass[review,number,sort&compress]{elsarticle}
\usepackage[english]{babel}
\usepackage[utf8x]{inputenc}
\usepackage{amsmath}
\usepackage{amssymb}
\usepackage{wasysym}
\usepackage{graphicx}
\usepackage{ulem}
\usepackage{lineno}
\usepackage{hyperref}
\usepackage[shortlabels]{enumitem}
\usepackage{setspace}
\usepackage{verbatim}

\begin{document}
\begin{frontmatter}
\title{Particle detection at cryogenic temperatures with undoped CsI}

\author[qu]{M. Clark\corref{cor1}}
\ead{m.clark@owl.phy.queensu.ca}
\author[qu]{P. Nadeau}
\author[qu]{S. Hills}
\author[ly]{C. Dujardin}
\author[qu]{P. C. F. Di Stefano}

\cortext[cor1]{Corresponding author}
\address[qu]{Department of Physics, Engineering Physics \& Astronomy, Queen's University, Kingston, Ontario, Canada, K7L 3N6}
\address[ly]{Institut Lumi\`ere Mati\`ere UMR 5306, Universit\'e Lyon, Universit\'e Claude Bernard Lyon 1, CNRS, F-69100, Villeurbanne, France}

\begin{abstract}
Scintillators are widely used as particle detectors in particle physics.
Scintillation at cryogenic temperatures can give rise to detectors with particle discrimination for rare-event searches such as dark matter detection.
We present time-resolved scintillation studies of Caesium Iodide (CsI) under excitation of both $\alpha$ and $\gamma$ particles over a long acquisition window of 1~ms to fully capture the scintillation decay between room temperature and 4~K. 
This allows a measurement of the light yield independent of any shaping time of the pulse.
We find the light yield of CsI to increase up to two orders of magnitude from that of room temperature at cryogenic temperatures, and the ratio of $\alpha$ to $\gamma$ excitation to vary significantly, exceeding 1 over a range of temperatures between 10 and 100~K.
This property could be useful in separating $\alpha$ backgrounds from the low energy nuclear recoil signal region.
We also find the time structure of the emitted light to follow similar exponential decay time constants between $\alpha$ and $\gamma$ excitation, with the temperature behaviour consistent with a model of self-trapped exciton de-excitation.  
Based on these properties, undoped CsI is an interesting candidate for use in cryogenic particle detectors.

\end{abstract}


\end{frontmatter}

\section{Introduction}
Solid-state scintillators are widely used as particle detectors at room temperature, in fields as diverse as national security, medicine, and particle physics.  
The light they emit when a particle interacts in them is a proxy for the energy deposited by the particle.  
Compared to other types of particle detectors, scintillators offer a wide range of target nuclei, and for some materials,  very fast response.  
In addition, in certain cases, the quantity of emitted light (light yield, or LY) or its timing, can also provide information on the nature of the interacting particle.  
CsI, pure or doped, is a frequently used scintillator, for instance in  neutrino physics~\cite{Akimov:2017}, or in searches for hypothetical dark matter particles~\cite{Kim:2012rza} that could account for most of the matter in our Universe~\cite{schnee_introduction_2011}.
There has been recent interest in the use of CsI as a cryogenic scintillating calorimeter for the detection of direct dark matter interactions with regular matter~\cite{Derenzo:2016fse,mikhailik_2014,cerdeno_scintillating_2014,Nadeau:2014kta,Zhang:2016}.  
CsI is an appealing target for such a detector because of its high LY at low temperature and the possibility of probing new WIMP interaction parameter space~\cite{cerdeno_scintillating_2014}. 
As an alkali halide material, it could also help to shed some light on the long-standing DAMA/LIBRA detection claim~\cite{Bernabei:2008ec,Bernabei:2013ax}, by performing a dark matter experiment with a similar material.

In comparison to the simple scintillation detectors used by DAMA, which are
incapable of event-by-event background discrimination, scintillating calorimeters can provide more insight into the nature of the interacting particle. 
A scintillating calorimeter is a particle detector consisting of a scintillating crystal held at cryogenic temperatures ($\lesssim$~50~mK) as the target medium, read-out by a light detector and thermal sensors, such that, for a given particle interaction, both scintillation and phonon signals can be observed~\cite{schnee_introduction_2011}. 
The phonon signal is a good proxy for the energy deposited by the particle.
For a given energy deposit, ionizing radiation in the form of alphas ($\alpha$), gammas ($\gamma$), and betas ($\beta$) will produce different amounts of scintillation light compared to the nuclear recoils expected to be caused by dark matter particles and neutrons, and possibly compared to one another.
This process, known as quenching,  allows for very powerful discrimination against background events. The CRESST experiment, for example, uses an array of scintillating calorimeters to detect nuclear recoils from dark matter particles~\cite{Angloher:2012tx}.
The possible sensitivity of such an alkali halide detector has been explored in simulations~\cite{nadeau_cryogenic_2015} and by the COSINUS experiment~\cite{Angloher:2016CsI}, which is planning on using cryogenic undoped NaI to evaluate the DAMA/LIBRA claim~\cite{Angloher:2016}. 
This approach is complementary to larger, room-temperature searches, 
including 
ANAIS~\cite{amare_preliminary_2014},
COSINE~\cite{Adhikari:2017,Cherwinka:2014xta},
PICO-LON~\cite{Fushimi:2016},
and SABRE~\cite{Xu:2014}.

The scintillation of CsI (with and without doping) has been studied over a range of temperatures~\cite{lamatsch1971kinetics,Kubota:1988,Schotanus:1990tk,Nishimura:1995,Amsler:2002kq,Moszynski:2005hu,mikhailik_2014}. 
The LY tends to increase as the temperature of the scintillator decreases. 
The ratio between $\alpha$ and $\gamma$ light emission is an interesting value to probe as it can allow for particle discrimination between $\alpha$ and $\gamma$ interactions.  
The difference in the ionization density between these two particles can affect the portion of deposited energy that is converted to light in the scintillator.

In the following, we report on our time-resolved measurements of the light yield of CsI between 300--3.4~K, under both $\alpha$ and $\gamma$ excitations for the first time, and over an unexplored time window~\cite{nadeau_cryogenic_2015}.
We use a novel zero-suppression data-acquisition technique enabling a large acquisition window of 1 millisecond to fully resolve the decays and reduce bias from losing light from the end of the window.
Our simultaneous measurement of $\alpha$ and $\gamma$ light from undoped CsI  provides a determination of the ratio of light emission between $\alpha$ and $\gamma$ excitations, hereafter referred to as the $\alpha/\gamma$ ratio ($R_{\alpha/\gamma}$), over this temperature range.
Our measurements are motivated by the possible use of CsI in rare-event searches at cryogenic temperatures.

\section{Experimental setup and sample preparation}
An optical cryostat produced by ColdEdge Technologies\footnote{Cold Edge Technologies, Allentown PA, www.coldedgetech.com} is used to cool the samples to any temperature down to 3.4~K, as has been described in previous publications~\cite{verdier_2.8_2009,Verdier:2011pb,Verdier:2012ie,Veloce:2015slj}. 
The crystal sample is mounted inside the cryostat and excited by $\alpha$ particles and $\gamma$ 60~keV quanta from an internal collimated $^{241}$Am source mounted to the sample holder. An external $^{57}$Co source can also be used to test $\gamma$ energies of 122~keV.
The $^{241}$Am $\alpha$ source used in this work was adapted from a common smoke detector where a protective film covers the radioactive material.  Using a silicon detector, we have measured the degraded mean energy of the $\alpha$ particles coming from the source to be 4.7~MeV~\cite{vonSivers:2015}.

\begin{figure}
	\centering 
		\includegraphics[width=.5\columnwidth]{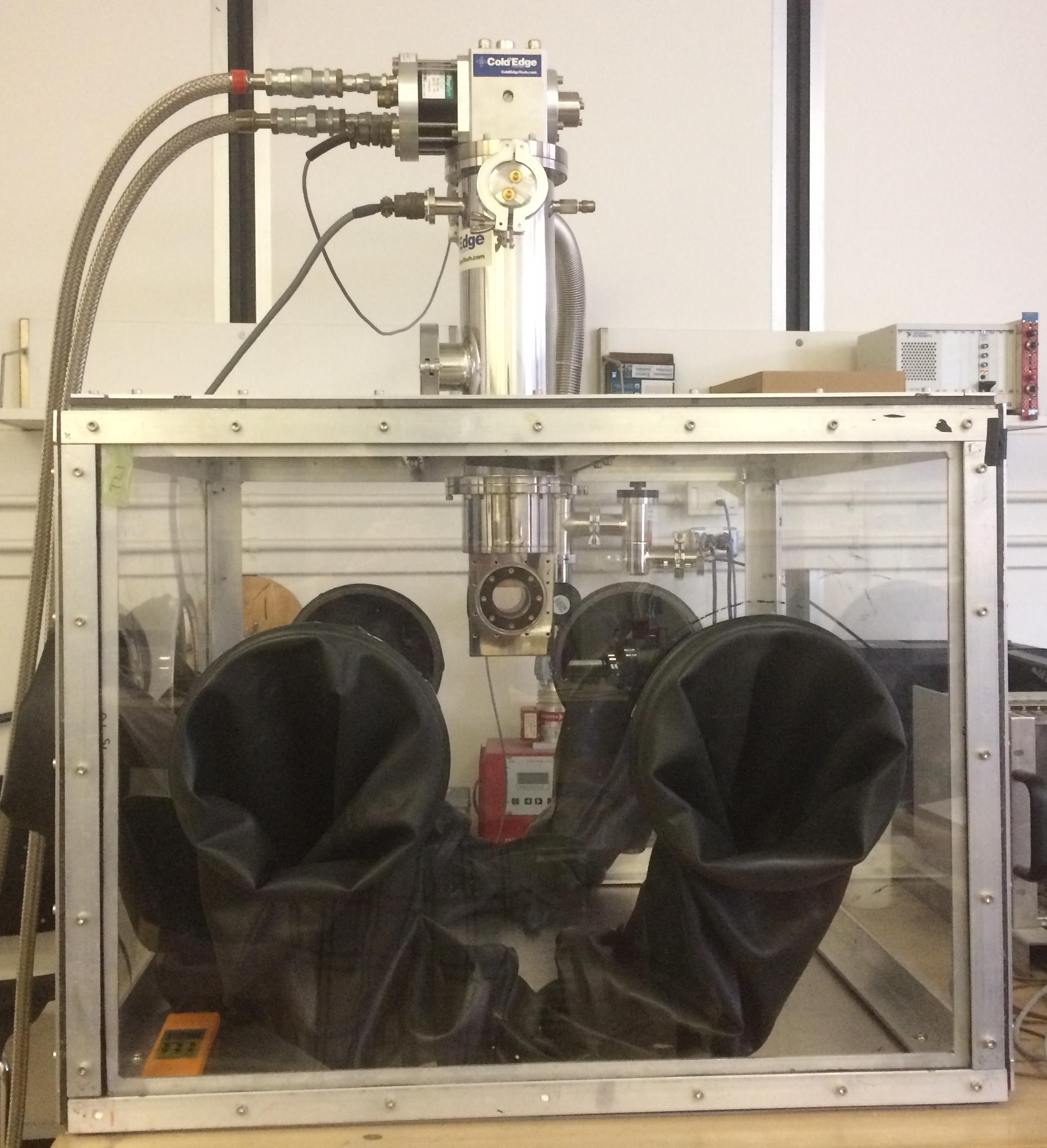}
		\caption{\label{fig:setuppic}The optical cryostat and glove box.}
\end{figure}

In order to facilitate our study of hygroscopic crystals we have built a glove box around the cryostat that was constantly flushed with compressed ``extra dry'' air ($<10$~ppm H$_2$O) and contained several bags of desiccant.
A photograph of the cryostat and glove box setup is shown in Figure~\ref{fig:setuppic}.
The nominally pure CsI crystal samples studied in this work were purchased from Hilger Crystals\footnote{www.hilgercrystals.co.uk} and have square cuboid geometries with dimensions $5\times5\times2$~mm$^3$ and all sides polished. 
The samples were adhered to a custom-made sample holder using silver paint instead of being mechanically held in place since the alkali halides are slightly brittle. 
We determined the optimal amount of adhesive to avoid cracking the crystal through trial and error on other samples, as there is differential contraction between the sample and holder during thermal cycling. 
The sample and holder were then moved into the glove box, installed onto the cold finger of the cryostat, and put under vacuum as quickly as possible. 
No changes in the optical quality of the sample were observed by eye during this procedure.

\begin{figure}
	\centering 
		\includegraphics[width=\columnwidth]{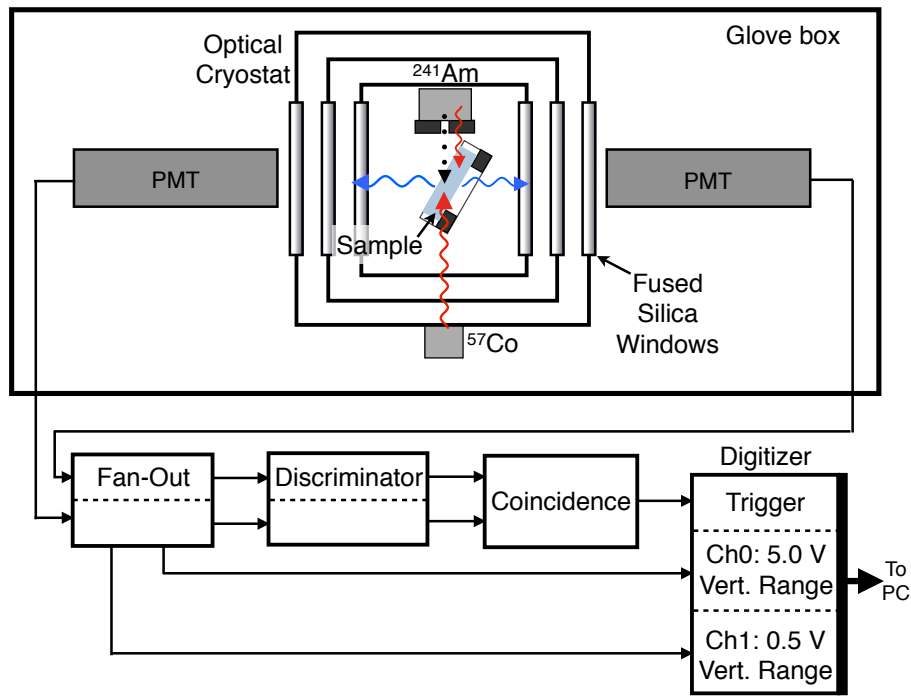}
		\caption{\label{fig:setup}A schematic of the setup for scintillation studies. A $5\times5\times2$~mm$^3$  sample is secured at the centre of the cryostat optical chamber on a holder, and is observed by two photomultiplier tubes (PMT). The volume of the optical cryostat is held at vacuum throughout the experiment whereas the glove box is filled with dessicated air.
The measurement of $\alpha$ particles (dotted arrow) and $\gamma$ quanta (incoming wavy arrows) is achieved using an internally mounted $^{241}$Am source. Optionally, other $\gamma$ sources can be mounted to the outside of the cryostat. The coincidence of the two PMTs provides a trigger for a digitizer that records photon signals from the PMT that is in view of the completely uncovered side of the crystal (the other PMT is partially masked by the holder).}
\end{figure}

As illustrated in Fig.~\ref{fig:setup}, the crystal sample was at an angle of {30\textdegree} with respect to the $^{241}$Am source so that the emitted $\alpha$ particles are incident with one of the smooth, large faces of the sample.  
At the same time, this face is well exposed to a 28~mm diameter PMT (Hamamatsu R7056) through a fused quartz window, with a second exposed to the rear face, partially obstructed by the sample holder. 
These PMTs have sensitivity to wavelengths down to 185~nm (the emission spectrum of CsI \cite{Kubota:1988} is known to have UV components) and a maximum quantum efficiency of 25\% at 420~nm. 
The PMT that is in view of the unobstructed face of the sample is used as the primary PMT for data acquisition, while the other is used to assist triggering. 
Based on the solid angle of the crystal exposed to the primary PMT and transmittance of the windows, we expect a light collection efficiency of roughly $10\%$~\cite{nadeau_cryogenic_2015}.
Note that no optical filters were used, so all light within the sensitivity of the PMT contributed to the light yield, as is standard in particle detection.

\section{Data acquisition}
\label{sec:data_acq}
Data acquisition is based on the multiple photon counting coincidence technique (MPCC), where the trigger for acquisition is produced when both PMTs detect photons within a given coincidence time window, allowing the measurement of both the light yield and the decay time constants of a scintillator \cite{Kraus:2005vj}. 
We have previously adapted this technique to the study of bismuth germanate (BGO) \cite{Verdier:2011pb} and ZnWO$_4$ \cite{Verdier:2012ie,di_stefano_counting_2013}. 
Signals from the PMTs pass through a series of Phillips Scientific Nuclear Instrumentation Modules (NIM) to produce a trigger pulse for the digitizer: first, a fan-out module to copy the signal; then, a discriminator to discard any low amplitude events below a given threshold, producing a flat logic pulse with length equal to the desired coincidence time window; and finally, a logic module that performs an AND operation on the incoming logic pulses on both channels and accepts any events where the pulses overlap, thus creating the coincidence condition. 
The logic module provides the trigger pulse to an 8-bit National Instruments digitizer (PXI-5154), which then records the incoming pulses from the PMTs at a 1~GHz sampling rate.
The full setup is described in Fig.~\ref{fig:setup}.

The coincidence window was set to 30~ns and an acquisition window of 1~ms (50~$\mu$s pre-trigger and 950~$\mu$s post-trigger) was used for all events at all temperatures. 
A long acquisition window was chosen so that we are sure that the full scintillation pulse could be observed at all temperatures.
To deal with the large amounts of data, a custom-made LabVIEW interface records only samples of the incoming signals that exceed a certain threshold to save hard-disk space and analysis time.
This introduces a light-yield and time-constant dependent bias on the estimation of the light yield which has been corrected.  This correction is  described in detail in Section~\ref{sec:data_anal}.
This is particularly important to ensure an unbiased comparison between $\alpha$ and $\gamma$ interactions.

The scintillation pulses from CsI have been characterized to have at least one very prominent fast decay time ($\sim$10-100~ns) \cite{Kubota:1988,Schotanus:1990tk} and other slower decay times ($>$1~$\mu$s)~\cite{Schotanus:1990tk}. 
At the start of a high light-yield scintillation event, many photons arrive at the PMTs faster than can be resolved individually; as a result, the PMTs output a high amplitude spike several $\mu$s wide (``pile-up'' or analog regime). 
By contrast, the slow component is slow enough such that single photons can be resolved easily by the PMTs (counting regime).  
Due to the vertical resolution of our digitizer, we therefore cannot use the technique we have developed based on a streaming time-to-digital converter to measure these pulses, since it depends on being able to resolve individual photons~\cite{di_stefano_counting_2013}. 
In order to measure the full pulse (fast $+$ slow components) with our digitizer, two copies of the signal from the primary PMT (facing the unobstructed side of the sample) are obtained from the Fan-Out module, as shown in Figure~\ref{fig:setup}. 
The signals are then sent to separate channels on the digitizer, where one channel is set to the maximum vertical range (5~V) so that the full amplitude of the initial fast spike can be measured without saturation, and the other channel is set to a low vertical range (0.5~V) in order to resolve the slow, individual photoelectrons that follow.

\begin{figure}
    \centering
	\includegraphics[width=\columnwidth]{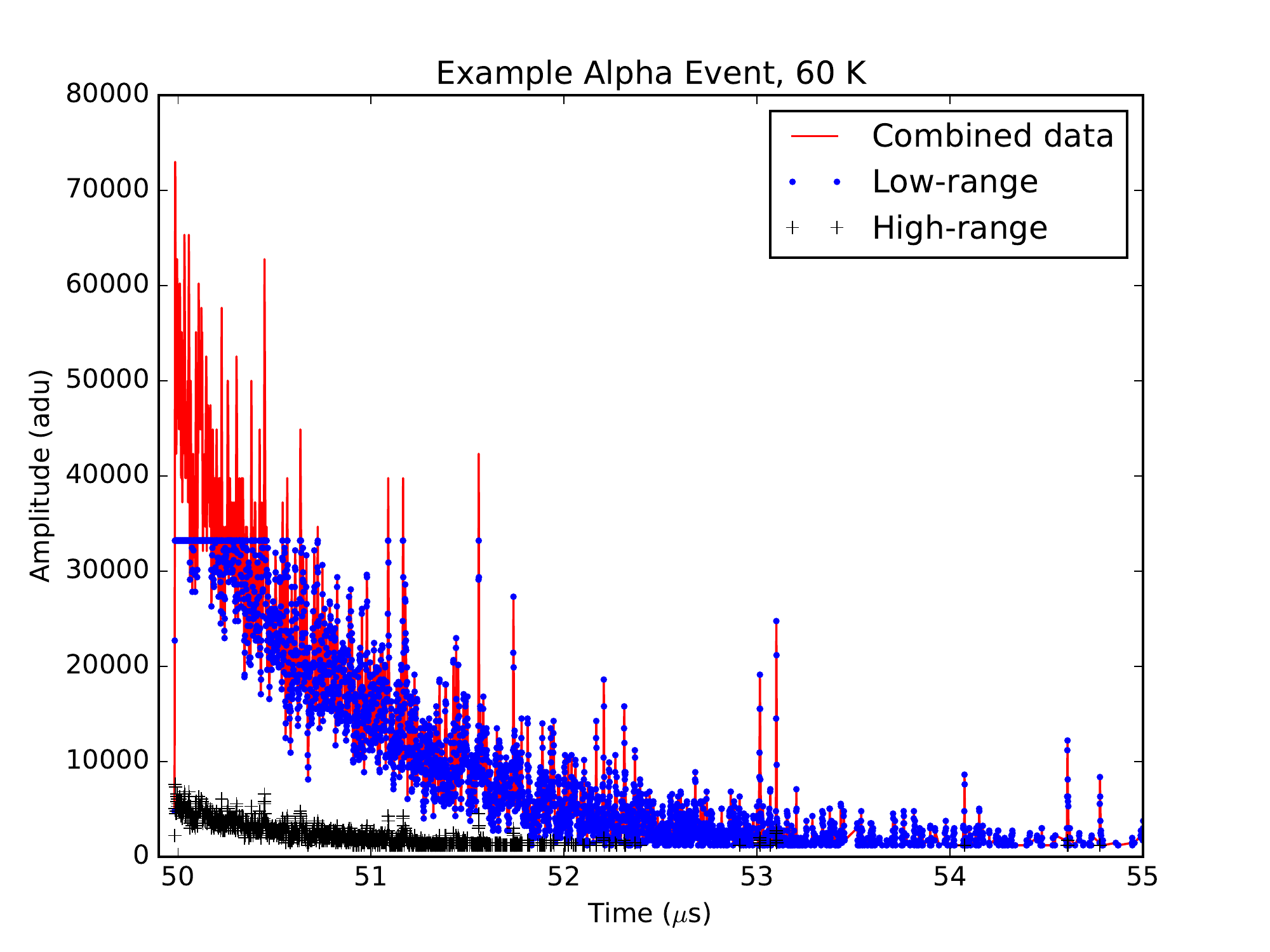}
	\caption{\label{fig:pulse} Demonstration of pulse reconstruction. Two channels with different ranges are used to record the same pulse, and are later combined through software (red line).  Saturation in the low-range channel (blue dot) is fixed using properly scaled values from the high-range channel (black cross), and pulses below precision in the high-range are still caught in the low-range channel. Pulse begins after a 50~$\mu$s pre-trigger and continues until 1~ms, though only the first 5~$\mu$s are shown here.}
\end{figure}

Data from both digitizer channels are recorded and then combined off-line to reconstruct the full pulse by replacing all of the saturated samples on the low range channel with their corresponding unsaturated samples on the high range channel after applying the scaling factor and offset. 
This results in a single channel of data with complete, unsaturated, reconstructed pulses with a larger dynamic range than possible with a single channel. 
An example of a single pulse reconstruction is shown in Figure~\ref{fig:pulse}. 
The reconstructed data are saved in an ASCII file as arrays of time and amplitude for each digitizer sample above the LabVIEW threshold in each triggered scintillation event.

\section{Data analysis methods}
\label{sec:data_anal}
The standard MPCC technique determines the light yield and the decay time components based on the number of photons in a given event and the known arrival time of each photon~\cite{Kraus:2005vj}. 
A set of cuts are applied to the data in order to reject spurious events such as events where multiple scintillations  occur in the acquisition window, and events where photons from a previous event straggle into the pre-trigger. 
Events with more than a single scintillation will have more photons in the later part of the time window, and thus can be identified by comparing the mean arrival time of the photons in a given scintillation event to the distribution of the mean arrival time of all the events. 
Events with photons in the pre-trigger region can be rejected using the distribution of the first photon in each event.

\begin{figure}
    \centering
	\includegraphics[width=\columnwidth]{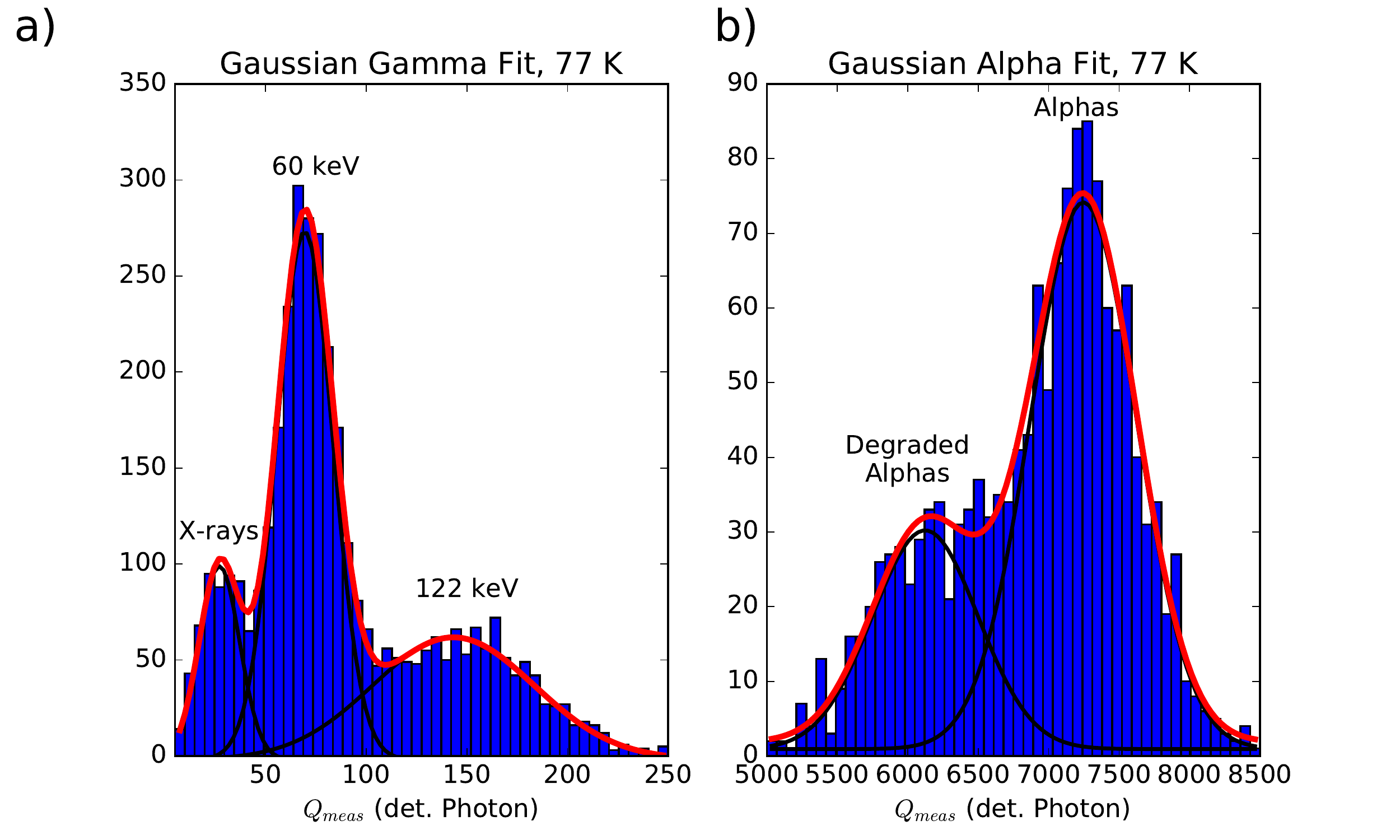}
	\caption{\label{fig:spectrum}Example of a spectrum of detected photons, here obtained at 77~K from the $\alpha$s,  $\gamma$s  and X-rays of $^{241}$Am and $^{57}$Co.}
\end{figure}

In the standard MPCC technique, the spectrum from a given source is obtained from the histogram of the number of photons in each event; after cutting the spectrum on a particle type and energy, the histogram of all photons with respect to their arrival time gives an average pulse shape~\cite{Kraus:2005vj}. 
We have adapted this technique to account for the early burst of unresolvable photons in the alkali halide pulses by building a histogram of the total charge in each event, instead of the arrival time of each photon, to produce the spectra shown in Figure~\ref{fig:spectrum}. 
A fit to the peak in the spectrum from a given particle at a given energy provides a measurement of the light yield (LY) of the scintillator as described in Section~\ref{sec:data_anal_ly}.
A secondary cut is applied to the data set on the total charge to separate the events that are a result of $\alpha$ or $\gamma$ interactions.
We use $\pm$ 1-$\sigma$ from the mean of each spectrum to determine where these cuts will be placed.

We can also reconstruct the average pulse for an $\alpha$ or $\gamma$ interaction by combining the charge vs. time of all pulses that are within the corresponding spectrum. 
This allows us to determine the scintillation decay time constants separately for excitations from the different particles. 
The combined pulse is then fit with a series of exponentials to extract the expected exponential decay time constants.
This process is described in Section~\ref{sec:data_anal_ts}.

\subsection{Light Yield}
\label{sec:data_anal_ly}
Thanks to our cryostat and our new data treatment, we are simultaneously sensitive to both $\alpha$ and $\gamma$ interactions at all temperatures, even with their large energy difference as illustrated in Figure~\ref{fig:spectrum}.
It allows us to measure charge over a large spread of particles (60~keV--4700~keV) and different temperatures.
The charge detected for a given particle varies by two orders of magnitude over the temperature range.

We note that at temperatures for which the LY enables sufficient resolution, an asymmetry in the $\alpha$ line becomes evident as a secondary peak that appears 15\% lower than the main peak.
We have verified with a Si detector that it does not come from the source, and we attribute the shape to small scratches on the polished surface, visible under microscope, changing the energy deposition of the $\alpha$ particles, which interact primarily with the surface of the crystal (interaction depth 24~$\mu$m~\cite{ASTAR}).  
The $\gamma$ LY is unaffected by the surface condition, as 60~keV $\gamma$s have a mean free path of 280~$\mu$m~\cite{XCOM}
We have therefore modeled the $\alpha$ line by two gaussians with fixed ratios of amplitudes, means and widths based on a free fit at the temperature with the greatest light yield (60~K).
For the $\alpha$-event selection cuts, we consider the higher peak to be representative of the $\alpha$ LY.

\begin{figure}
    \centering
	\includegraphics[width=\columnwidth]{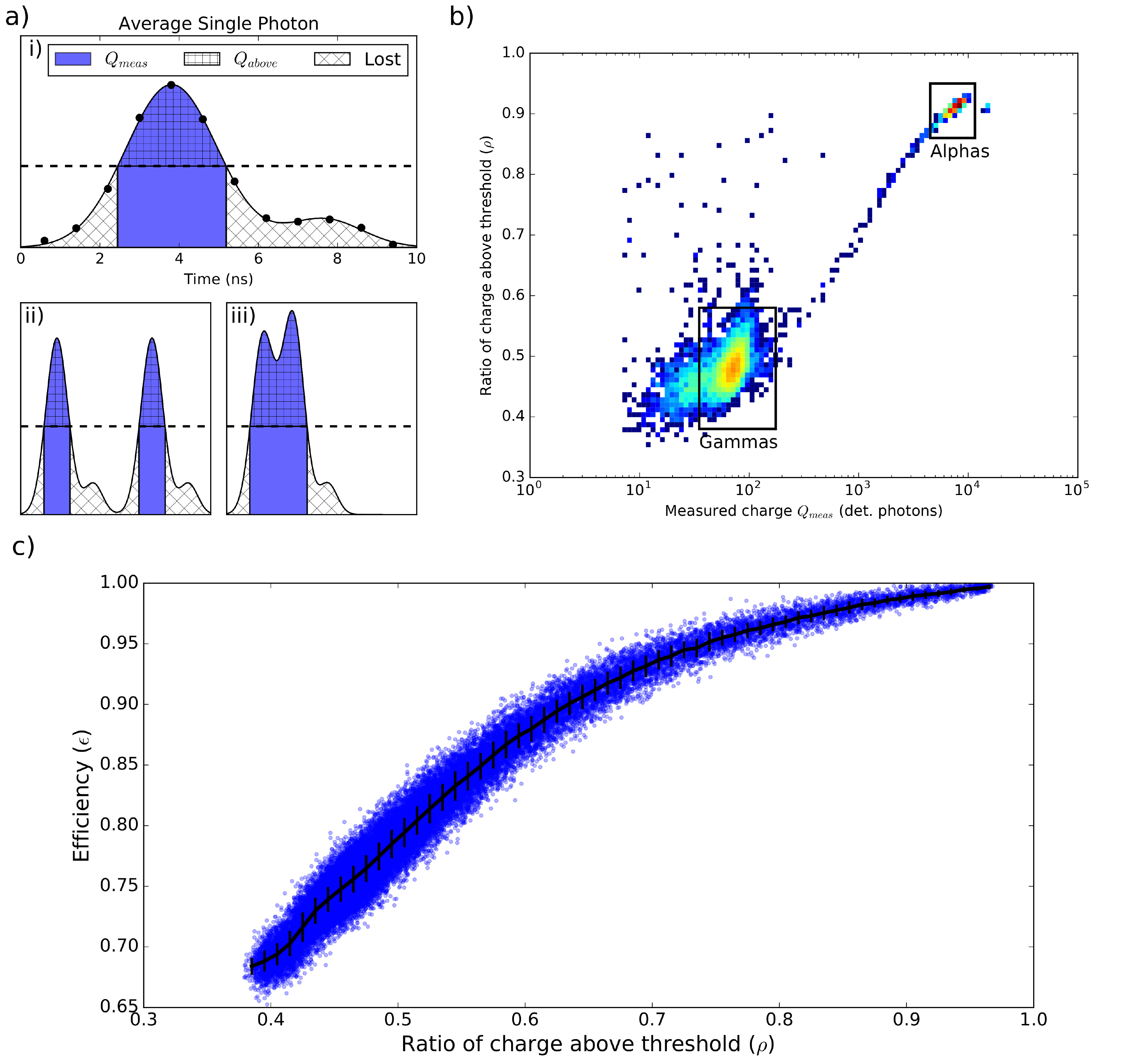}
	\caption{\label{fig:thresh}Effect of threshold on measured LY.  Fig.~\ref{fig:thresh}a-i shows that a portion of a single photoelectron charge is lost because the amplitudes are below the detection threshold.  
    $Q_{meas}$, shaded in blue, is the amount of charge that we record for a given event.
    $Q_{above}$, vertical cross-hatched, is the area under the curve but above the threshold. 
    This allows us to create the variable $\rho=Q_{above}/Q_{meas}$, the ratio of charge above threshold, to estimate the fraction of lost charge, as described in Section~\ref{sec:data_anal_ly}.
    If multiple photons are resolved in time as in Fig.~\ref{fig:thresh}a-ii, the fraction of lost charge is the same as for the single photoelectron.  
    If there is pileup of the photoelectrons as in Fig.~\ref{fig:thresh}a-iii, the fraction of lost charge is smaller. 
    At a given temperature, the data presented in Fig.~\ref{fig:thresh}b of $\rho$ vs. $Q_{meas}$ show that the effect is detrimental to the pulses from $\gamma$s compared to $\alpha$s, since the former emit fewer photons.  
    Fig.~\ref{fig:thresh}c shows simulations (blue spots) allowing to reconstruct the actual LY from the measured one using the median (black line); error bars are taken from spread of points.}
\end{figure}

As mentioned in Section~\ref{sec:data_acq}, we use an on-line threshold to reduce the amount of data that we are required to read and save to disk by excluding all voltage data below the threshold. 
We found, however, that doing so affected our measurement efficiency differently for different LYs.
Figure~\ref{fig:thresh}a-i shows the single photon response of the PMTs used, determined from a dedicated experiment without the use of the threshold.
When the threshold is applied, we lose a small amount of charge on each side of the distribution.
If photons arrive sufficiently spaced out in time as in Figure~\ref{fig:thresh}a-ii, the same portion of the response will be lost for each photon.
However, if several photons arrive in quick succession the responses can pile-up on each other, allowing the charge to exceed the threshold and be measured (as shown in Figure~\ref{fig:thresh}a-iii), which would increase the efficiency of that measurement.
The likelihood that photons will pile-up on each other depends on both the LY of the interaction as well as the decay constant of the scintillation, both of which change as a function of temperature. 
We correct for this effect for each particle interaction and each temperature point to be able to globally relate the LY of these interactions.
 
We define $Q_{total}$ as the real total amount of charge that the scintillation event has induced in the PMT. 
This is the value that we want to have access to in order to compare the LY under $\alpha$s and $\gamma$s at all temperatures.
$Q_{meas}$ is defined as our actual measured charge.
The efficiency of a measurement $\epsilon = Q_{meas}/Q_{total}$ will change at each temperature and for each type of particle interaction. 
Since $Q_{total}$ is inaccessible, we define $\rho=Q_{above}/Q_{meas}$, the ratio of the integral above the threshold $Q_{above}$ to the total measured integral $Q_{meas}$, as a proxy for the efficiency $\epsilon$.
In Figure~\ref{fig:thresh}b, we plot the variable $\rho$ against $Q_{meas}$ which is representative of the emitted light, for an example temperature. 
$\alpha$ interactions can be seen in the top right of the plot, with high emitted light and high $\rho$. 
$\gamma$ interactions are seen in the middle of the plot, with a lower $\rho$ and lower emitted light. 
From this plot, we can see that the efficiencies for $\alpha$ and $\gamma$ interactions are systematically different because of the very different LY causing differing amounts of pile-up between photons.

We carried out simulations to relate $\rho$ to the efficiency $\epsilon$. 
In the simulations, we used different numbers of photons distributed in time by an exponential distribution with varying decay time constants to produce different values of $\rho$. 
Determining the relationship between these two variables in simulated data allows us to make an estimate on the efficiency in our recorded data. 
The relationship between $\rho$ and $\epsilon$ is shown in Figure~\ref{fig:thresh}c, allowing us to make corrections to our LY and ratio $R_{\alpha/\gamma}$. 

\subsection{Time Structure}
\label{sec:data_anal_ts}

Our data acquisition system is able to sample the charge induced in the PMTs at a sampling rate of 1~ns.
Using this time-resolved data, we are able to create an ``average pulse'' by combining the charge vs. time data for a set of individual traces that correspond to the $\alpha$ or $\gamma$ population.
We chose to use the coincidence trigger time as the beginning of each pulse; because we are only recording the signal from one PMT, and the trigger is set by a 30~ns coincidence window, the actual first pulse recorded may be anywhere from 0-30~ns before the trigger time.
Thus, we do not expect to be sensitive to time structure below this order of magnitude.

For each population, the average pulse histogram is fit to five exponential decays plus a constant background to model the scintillation decay and background noise.  
We use the same functional fit for all temperatures to fit all pulses in the same unbiased way.
We expect at least three exponential decay components~\cite{mikhailik_2014} at low temperature, and additional exponentials allow for a good fit of extra-slow components, though we don't attempt to interpret them.

\begin{figure}
    \centering
	\includegraphics[width=\columnwidth]{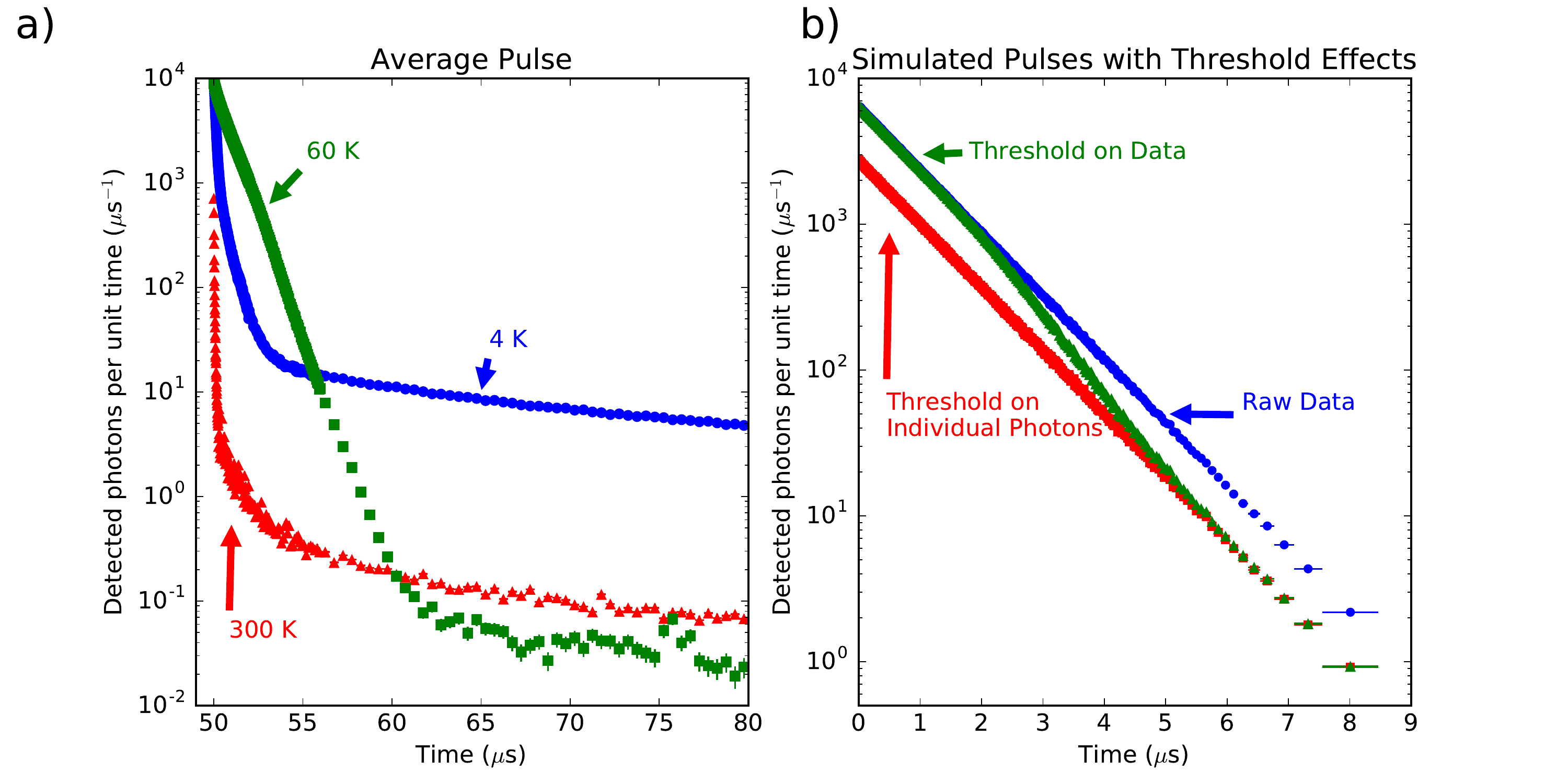}
	\caption{\label{fig:avepulse}a) Example of average $\alpha$ pulses for different temperatures (note irregular binning).  At least two exponential decays are evident for each temperature, spanning nearly 6 orders of magnitude in intensity.  Kink in pulse, caused by variable charge-reconstruction efficiency, is visible at 60~K between 52--56~$\mu$s.
     b) Simulations of the kink (Sec.~\ref{sec:data_anal_ts}). Simulated data for a single time constant with no threshold applied (blue circles); the threshold applied to the total pulse, with photons piling up on each other, as done in the real data (green triangles); and the threshold applied to every arriving photon as if they arrived separately (red squares). Kink occurs in transition between regions where photoelectrons pile up and where they don't, but does not affect time constant on either side of transition. }
\end{figure}

As mentioned in Section~\ref{sec:data_acq}, the application of a threshold changes the efficiency of detection for individual photons versus pile-up photons. 
This has an effect on the average pulse shape separate from the measurement of LY.
The effect can be seen in Figure~\ref{fig:avepulse}a as a kink in the pulse, at a time where the likelihood of a photon arriving at the PMT decreases to a point where pile-up is no longer likely (see Figure~\ref{fig:thresh}).  
We have confirmed through simulation, shown in Figure~\ref{fig:avepulse}b, that the change in efficiency due to photons piling-up only impacts the pulse in this transition region.  
Before and after, the shape of the pulse follows the underlying distribution.
To deal with this in the fit of exponentials, the kink time section has been excluded, and the fit keeps the same time constant before and after the kink time.  
Lastly, since the pulse shape spans multiple orders of magnitude, non-uniform binning has been utilized to reduce bin-to-bin statistical fluctuations in the late portion of the pulse.

\section{Results and discussions}
\label{sec:disc}

\subsection{Light yield and $\alpha/\gamma$ ratio}
\label{sec:disc:LYQ}
The temperature response curves for the LY of CsI under $\alpha$ and $\gamma$ excitation were measured for several stabilized temperatures while cooling from 300~K to 3.4~K.
Our method allows us to measure LY without the use of a shaping time, making the measurement independent of any changes in decay time constant vs. temperature.
As a proxy for the number of emitted photons, we take the number of photons detected in the front PMT.
This number of photons is determined from the charge detected by the PMT thanks to a separate measurement with a low-light source providing the average charge induced in the PMT by a single photon.

Ideally, these results would take into account possible temperature-dependent shifts of the emission spectra relative to the fixed quantum efficiency of the PMTs.
Comparing emission spectra at low temperatures with the quantum efficiency of our PMT, we estimate that, relative to room temperature, our detection efficiency changes by at most 10\% over the entire temperature range~\cite{Nishimura:1995,mikhailik_2014}.

\begin{figure}
   \begin{center}
  \includegraphics[width=0.6\columnwidth]{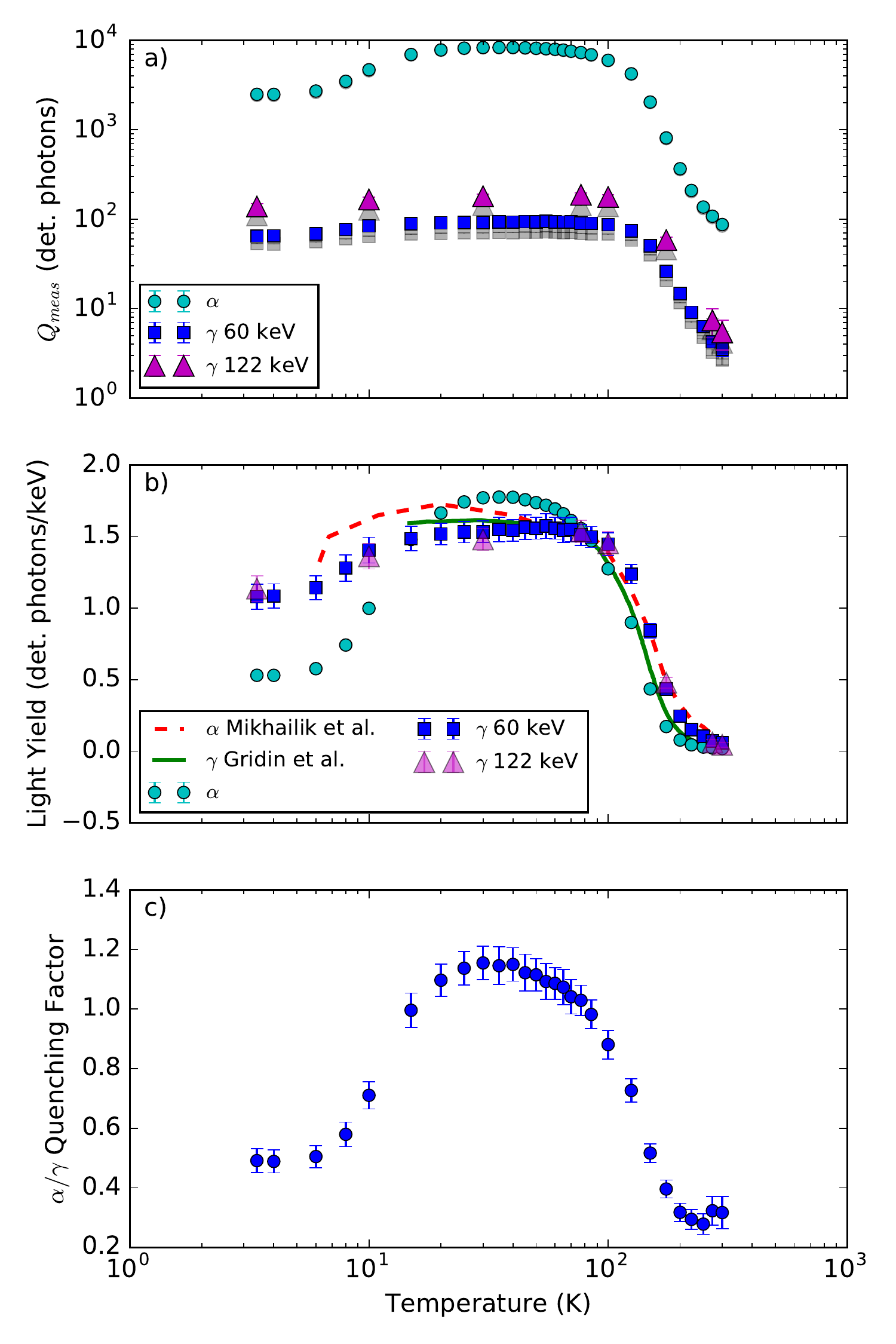}
  \end{center}
  \caption{\label{fig:LYQF}a) Detected photons as a function of temperature, for 4.7~MeV $\alpha$s (circles) and $\gamma$s (60~keV: squares; 122~keV: triangles). A correction has been applied (original: light ; corrected: solid) to compensate for the loss in charge detected from our threshold (see Sec.~\ref{sec:data_anal} for details).
  b) Detected photons (corrected) per unit deposited energy as a function of temperature for $\alpha$s and $\gamma$s, showing results for 60~keV and 122~keV $\gamma$s are consistent within errorbars. Results of $\alpha$-excitation from Mikhailik et al.~\cite{mikhailik_2014}  are shown by the dashed red and of $\gamma$-excitation from Gridin et al.~\cite{Gridin:2015} are shown by the solid green line, scaled to be equal to our result at 77~K.
  c) $\alpha$/$\gamma$ ratio (corrected) as a function of temperature, showing a factor 4 variation and reaching values greater than 1.  In a) and b), conversion of detected photons to emitted ones requires a multiplication by roughly 70, though this does not affect ratio in c).} 
\end{figure}

The number of detected photons for the various particles and energies are shown as a function of temperature in Figure~\ref{fig:LYQF}.
Except for $\gamma$s at temperatures above 200~K, at least 10~photons are detected per event. 
The temperature dependence of the number of detected photons over energy, a proxy for LY,  is also shown in Figure~\ref{fig:LYQF}b. 
This LY has been corrected for efficiency as described in Section~\ref{sec:data_anal_ly}
We see the $\alpha$ LY increase by a factor of 100 from 300~K to 30~K, and decrease to a factor of 30 above 300~K at 3.4~K, our lowest temperature. 
The $\gamma$ LY for both 60~keV and 122~keV does not see as much of an increase as for the $\alpha$s, rising by a factor 20-30 above that at 300~K below 100~K.
The increase in LY is consistent between the two different $\gamma$ energies, within uncertainty.
Overall, the LY is relatively constant at temperatures below 7~K.  We assume it remains so at even lower temperatures.

Though comparison is difficult because of uncertainties in our light collection efficiency and quantum efficiency of the PMTs, previous results of the absolute LY of CsI by Amsler et al.~\cite{Amsler:2002kq} give $50\pm5$~ph/keV at 80~K under $\gamma$ excitation.
Another measurement under $\gamma$ excitation by Moszynski et al.~\cite{Moszynski:2003} gives $107\pm10$~ph/keV for undoped CsI at 77~K.
At 77~K, we detect $83\pm5$ photons from a 60~keV $\gamma$-excitation. 
Assuming an efficiency of $\sim 10 \%$ due to numerical aperture and reflection of windows~\cite{verdier_2.8_2009}, and a PMT quantum efficiency of $\sim 15\%$ at the mean emission of CsI, this gives a rough estimate of $90$~ph/keV for the LY , which is compatible with the result from Amsler et al.
More recently, Sch\"affner et al.~\cite{Schaffner:2012ei} quote an absolute detected energy of 7.1\% from undoped CsI at 10~mK.
Using the mean energy of an emitted photon from CsI (3.9~eV~\cite{nadeau_cryogenic_2015}), and taking the light collection efficiency of the CRESST type light detectors to be 31\%~\cite{Kiefer:2012va}, this would imply an absolute LY of $60$~ph/keV.
At our lowest temperature, using the same method as above, we estimate the LY of CsI to be roughly $65$~ph/keV at 3.4~K, which appears to be compatible with the result from Sch\"affner et al.
The agreement of estimated LY values with previous measurements indicates that the corrections described in Section~\ref{sec:data_anal_ly} are performing correctly.

To compare with recent results from Mikhailik et al.~\cite{mikhailik_2014} and Gridin et al.~\cite{Gridin:2015}, we have scaled their reported LY values to ours at 77~K. 
Although Gridin et al. do not report time-resolved LY, we believe that our results are comparable because of our long acquisition window capturing full scintillation events.
We see the same general trend of LY increase as previous measurements, but different relative LY at high temperatures; this could be due to a difference in incidental Tl impurities in the crystals.
The light emission from Tl can be seen in the room temperature spectra of Mikhailik et al., but disappears at lower temperature; this could increase the LY preferentially at high temperatures in a crystal with a larger amount of Tl.

As our setup allows the measurement of the lines from $\alpha$ particles and $\gamma$ quanta emitted by $^{241}$Am,  we can determine the $\alpha$/$\gamma$ ratio, $R_{\alpha/\gamma}$, defined by the ratio of LY for $\alpha$s over the LY for $\gamma$s at a given energy $E$:
\begin{equation}\label{eq:qf}R_{\alpha/\gamma} (E)  = \frac{{\rm LY}_{\alpha}(E)}{{\rm LY}_{\gamma}(E)} = \frac{\mu_\alpha (E) / E}{\mu_\gamma (E) /E}  = \frac{\mu_\alpha (E) }{\mu_\gamma (E)} ,
\end{equation}
where $\mu_i (E)$ is the measured response to particles of type $i$ depositing energy $E$.
$R_{\alpha/\gamma}$ quantifies the relative difference in light produced by both types of interactions for an equivalent energy deposit, and thus is independent of our overall light collection efficiency barring effects related to the interaction depth of the particle.

The temperature dependence of $R_{\alpha/\gamma}$ is shown in Figure~\ref{fig:LYQF}c, as determined using the 60~keV $\gamma$ and 4.7~MeV degraded $\alpha$ from $^{241}$Am.
Remarkably in our data, $R_{\alpha/\gamma}$  becomes greater than 1 for temperatures between 10-100~K.
A value of $R_{\alpha/\gamma}$  greater than 1 is unexpected in general because of the higher ionization density of $\alpha$ particles leading to a higher probability of non-radiative recombination of electrons and holes before formation of self-trapped excitons~\cite{Sysoeva:1998}, though this behavior has previously been observed in alkali halide crystals grown by the Kyropoulos method~\cite{Birks:1964-11}, and in pure or doped ZnSe~\cite{Arnaboldi:2011ce,Sysoeva:1998,Klamra:2002bg,Nagorny:2017}.
Watts et al.~\cite{Watts:1962} observed an anomalous value of the $\alpha/\beta$ ratio (which should be equivalent to $R_{\alpha/\gamma}$) of $3.5\pm0.3$ at 77~K and 4~K for a single CsI crystal, that was not reproduced in other crystals, though they presented no explanation.

Birks suggests that a large temperature dependence in $R_{\alpha/\gamma}$ is associated with a high defect density, as the lower ionization density of $\gamma$-excitation allows electrons and holes to drift and become trapped in defects instead of forming self-trapped excitons~\cite{Birks:1964-11}.
There is evidence for this in ZnSe, where thermal treatment has been seen to affect $R_{\alpha/\gamma}$~\cite{Nagorny:2017}.
Another hypothesis is that there are significant excitation-dependent changes in the emission spectra, causing our detection efficiency to change between particles at various temperatures.
Lastly, there are three effects that could be considered, though they are unlikely because they would have to be strongly temperature dependent to explain the data:  the first being that the light collection efficiency differs significantly between the surface and the bulk of the sample, the second being that the scintillation properties of the surface differ from those of the bulk, and the third being the non-proportionality of the LY.

Since in practice our measurement involves $\alpha$ and $\gamma$ interactions of different energies (4.7~MeV and 60~keV respectively), the non-proportionality  ($NP$) of the scintillation response could also have an affect on our measured light ratio:
\begin{equation}
\label{eq:qf_e}
R_{\alpha/\gamma} (4.7 \mbox{ MeV}) =\frac{{\rm LY}_{\alpha}(4.7 \mbox{ MeV})}{{\rm LY}_{\gamma}(4.7 \mbox{ MeV})}= \underbrace{ \frac{{\rm LY}_{\gamma}(60 \mbox{ keV})}{{\rm LY}_{\gamma}(4.7 \mbox{ MeV})} }_{NP}  \underbrace{ \frac{{\rm LY}_{\alpha}(4.7 \mbox{ MeV})}{{\rm LY}_{\gamma}(60 \mbox{ keV})}}_{\text{measurement}}
\end{equation}
There is some evidence that the non-proportionality of the LY in pure CsI  depends on temperature.
Lu et al~\cite{Lu:2015} report that at 295~K, the LY of a 60~keV $\gamma$-interaction is greater  than that of a 1~MeV $\gamma$ by approximately 15\%, whereas at 100~K there is almost no difference.  
It is unlikely the observed changes in $R_{\alpha/\gamma}$ can be explained by this effect alone, however, as $R_{\alpha/\gamma}$ varies by a factor of 4 over the measured temperature range. 
Additionally, to be consistent with a value of $R_{\alpha/\gamma} < 1$ at all temperatures, the non-proportionality would have to decrease an additional 20\% between 100~K and 30~K. 
We are not aware of any experimental data at other temperatures for pure CsI, or for $\gamma$-interactions higher than $\sim 2$~MeV.

\subsection{Time Structure}
\label{sec:disc_TS}
\begin{figure}
    \centering	\includegraphics[width=0.9\columnwidth]{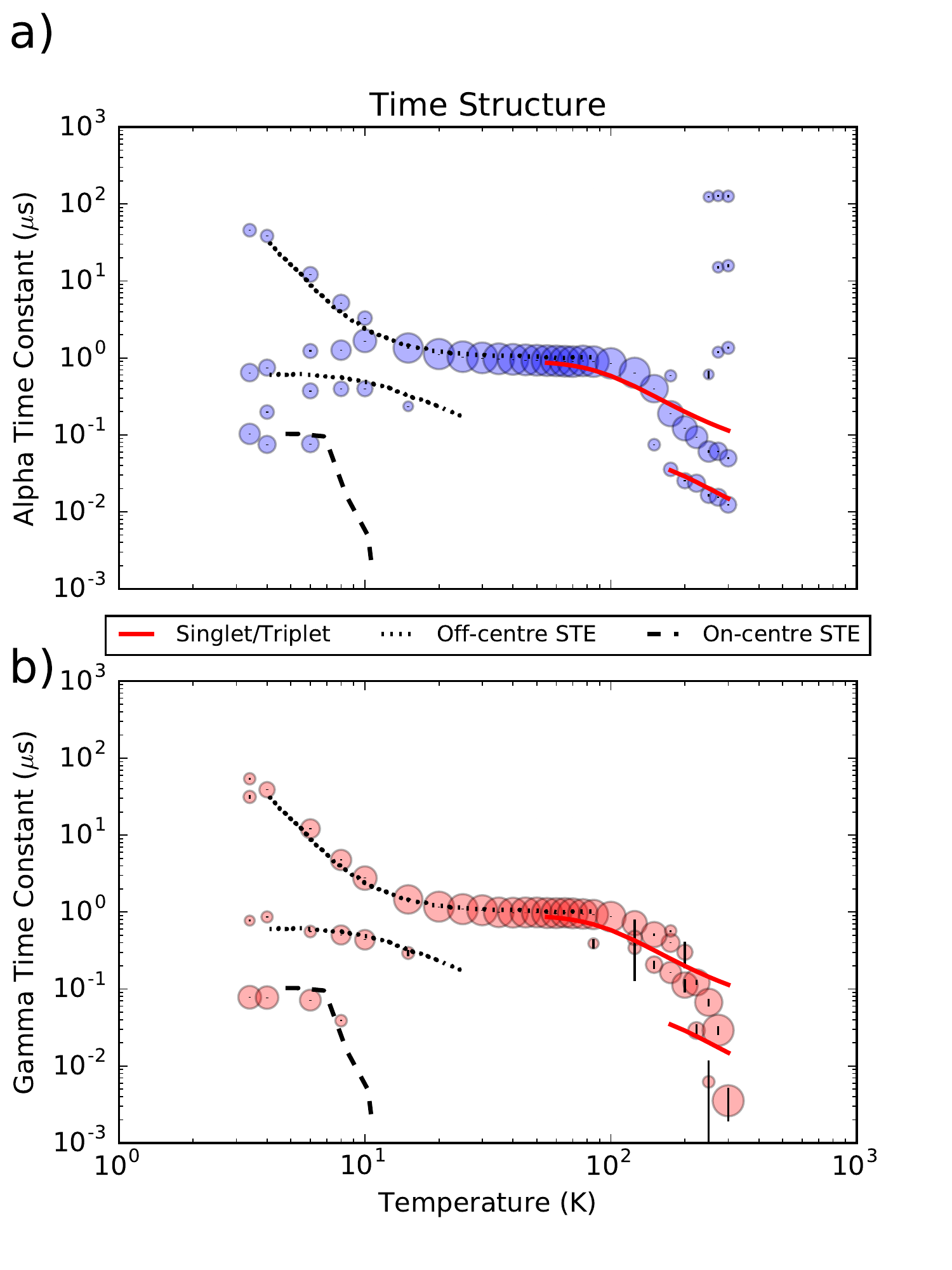}
	\caption{\label{fig:tau} 
	Evolution of time constants with temperature. At a given temperature, only time constants composing at least 10\% of LY are shown.  Marker size at a given temperature is proportional to the contribution of that time constant to the total LY at that temperature. Models of STE decay from Nishimura et al.~\cite{Nishimura:1995} are shown for the off-center STE state (dotted black line), on-centre STE state (dashed black line), and singlet and triplet STE state at high temperatures (solid red line).
    a) Time constants from $\alpha$ excitation 
    b) Time constants from $\gamma$ excitation
    }
\end{figure}

The results of the time constant fits mentioned in Section~\ref{sec:data_anal_ts} are summarized in Figure~\ref{fig:tau}.
Time constants from the fits that contribute at least 10\% of the total LY at that temperature are plotted in Figure~\ref{fig:tau}a for $\alpha$ excitation, and Figure~\ref{fig:tau}b for $\gamma$ excitation.
The area of the marker represents the contribution of that time constant to the total integral of the pulse, the largest points showing the most prominent time constant at a given temperature.

First, for $\alpha$ excitation, above 100~K there appears to be two prominent time constants, both shorter than 1~$\mu$s.
As the temperature decreases, the main time constant increases from 0.4~$\mu$s at 300~K to 1~$\mu$s at 100~K, where it becomes the only important time constant.
We also see a fast component that begins at 0.01~$\mu$s at 300~K, increasing in length to 0.1~$\mu$s at 150~K.
This behaviour is consistent with the measurements of Amsler et al.~\cite{Amsler:2002kq} from 100-300~K.

From 100~K to 10~K, there remains only one time constant at around 1~$\mu$s.
This component is consistent with previous measurements of undoped CsI scintillation at low temperature~\cite{mikhailik_2014,Watts:1962}
Below 10~K, the structure becomes more complicated, with several components contributing equal amounts of light to the pulse.
We see a short component of the order 100~ns from both $\alpha$s and $\gamma$s, consistent with the 290~nm emission,
and a long component that is consistent with the 338~nm emission, which are described by previous studies as a system of three excitonic absorption bands~\cite{lamatsch1971kinetics,Gridin:2015}.
There are several very long time constants present at high temperatures that seem to hold a constant value of 10~$\mu$s and 100~$\mu$s, which could be attributed to coincidence within our acquisition window of $\alpha$ and $\gamma$ events.
Alternatively, these could be attributed to the presence of shallow traps or defects.

For the $\gamma$ excitation, the main time constants show a very similar pattern to the $\alpha$ excitation data.
This indicates that the same light production mechanism is being excited by the two different radiation sources, which has also been seen previously with $\alpha$s and X-rays~\cite{mikhailik_2014}.
Long time constants visible by $\alpha$ excitation are absent in the $\gamma$ excitation data, which could indicate a larger concentration of defects near the surface of the crystal, as expected~\cite{Nishimura:1995}.

We have compared our results to the results of fits to a model of various STE decays by Nishimura et al.~\cite{Nishimura:1995} in Figure~\ref{fig:tau}. For both $\alpha$ and $\gamma$ excitation, the data corresponds well with the model at low temperatures. At high temperatures, two exponential components are seen in $\alpha$ excitation, corresponding to the singlet and triplet STE states~\cite{Nishimura:1995} seen as the solid red lines in Figure~\ref{fig:tau}.  At low temperature, the evolution of the two decay rates of the off-center STE, seen as the dotted black lines, follow the same trend in our data.  The on-center STE decay rate can also be seen as the dashed black line in Figure~\ref{fig:tau}, which we also see because of our lack of spectral discrimination between the two states.

\section{Conclusion}
We present the first measurement of the scintillation properties of undoped CsI over a wide range of temperature under both $\alpha$ and $\gamma$ excitation using a time-resolved zero-suppression measurement technique.
For the first time to our knowledge in CsI, a long measurement time window was used to capture the full scintillation event at each trigger, removing any bias from long or short scintillation times.

The LY under $\alpha$ excitation increases by nearly two orders of magnitude from 300--30~K, to a maximum value of roughly 120~ph/keV.
At the lowest measured temperature of 3.4~K, the LY remains a factor 30 above the level at room temperature.
Under $\gamma$ excitation, a factor 20 increase in LY is observed below 100~K when compared with the room temperature yield, with a maximum at roughly 90~ph/keV.
The ratio between $\alpha$ and $\gamma$ excitations fluctuates significantly over the temperature range, and was found to be greater than one for temperatures between 10--100~K, which could be an indicator of a large amount of defects in the crystal.
It also may be an artifact of particle specific emission changes as the temperature decreases.
Measurements of the time constants for both $\alpha$s and $\gamma$s agree with previous measurements, and follow previous models of STE decay~\cite{Nishimura:1995}.

These measurements further establish undoped CsI as an effective cryogenic scintillator with a high light yield at low temperatures. 
With such a light yield, above that of most other cryogenic scintillators~\cite{Mikhailik:2010}, low thresholds can be reached in experimental applications that require it, such as searches for dark matter.  In such searches, the high $\alpha$/$\gamma$ ratio at low temperatures indicates that the $\alpha$ background should remain separated from the nuclear recoil signal region due to the low nuclear recoil quenching factor of 0.1 for Cs and I~\cite{Angloher:2016CsI}.  More generally, light yield and $\alpha/\gamma$ ratio are high at liquid nitrogen temperatures, conditions that lend themselves well to practical applications.

\section{Acknowledgements}
This work has been funded in Canada by NSERC (Grant SAPIN 386432), CFI-LOF and ORF-SIF (project 24536).

\section{References}
\bibliographystyle{elsarticle-num}
\bibliography{mybib}

\end{document}